# Intelligent Emergency Message Broadcasting in VANET Using PSO

Ghassan Samara

Department of Computer Science, Zarqa University
Zarqa, Jordan

Tareq Alhmiedat

Department of Information Technology, Tabuk University
Tabuk, Saudi Arabia

Abstract— The new type of Mobile Ad hoc Network which is called Vehicular Ad hoc Networks (VANET) created a fertile environment for research.

In this research, a protocol Particle Swarm Optimization Contention Based Broadcast (PCBB) is proposed, for fast and effective dissemination of emergency messages within a geographical area to distribute the emergency message and achieve the safety system, this research will help the VANET system to achieve its safety goals in intelligent and efficient way.

Keywords- PSO; VANET; Message Broadcasting; Emergency System; Safety System.

## I. INTRODUCTION

Recent Year's rapid development in wireless communication networks has made Car to Car (C2C) and Car to Infrastructure Communications (C2I) possible in Mobile Ad hoc Networks (MANETs). This has given birth to a new type of high mobile MANET called Vehicular Ad hoc Networks (VANET) creating a fertile area of research aiming for road safety, efficient driving experience, and infotainment (Information and Entertainment) [1, and 2].

Creating a safety system on the road is a very important and critical concern for human today, each year nearly 1.3 million people die as a result of road traffic accidents – more than 3000 deaths each day - and more than half of these people are not travelling in a car, the injuries are about fifty times of this number [3]. The number of cars in 2004 is approximately estimated as 750 million cars around the world [4], with an annually constant increase by 50 million car around the world [5], with this constant raise, the estimated number of cars nowadays exceeding one billion, this raise the possibility to increase the number of crashes and deaths on the roads, road traffic accidents are predicted to become the fifth leading cause of death in the world, resulting in an estimated 2.4 million death each year as stated by the World Health Organization (WHO) [3], besides traffic congestion makes a huge waste of time and fuel, this makes developing an efficient safety system an urgent need on the road.

This research is funded by the Deanship of Research and Graduate Studies in Zarqa University/ Jordan.

The new techniques in this system should aim to make the intelligent vehicle to think, communicate with other vehicles and act to prevent hazards.

VANET safety applications depend on exchanging the safety information among vehicles (C2C communication) or between Vehicle to infrastructure (C2I Communication) using the control channel, see figure 1.

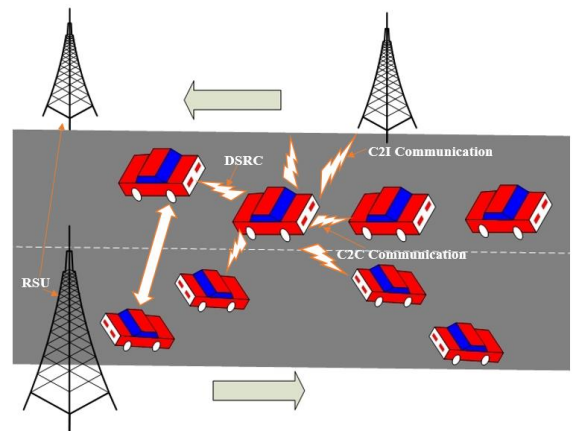

Figure 1: VANET Structure

VANET safety communication can be made by two means: Periodic Safety Message (called Beacon in this paper) and Event Driven Message (called Emergency Message in this paper), both sharing only one control channel. The Beacon messages are status messages containing status information about the sender vehicle like position, speed, heading …etc. Beacons provide fresh information about the sender vehicle to the surrounding vehicles in the network helping them to know the status of the current network and predict the movement of vehicles. Beacons are sent aggressively to neighboring vehicles 10 messages each second.

Emergency Messages are messages sent by a vehicle detect a potential dangerous situation on the road; this information should be disseminated to alarm other vehicles about a probable danger that could affect the incoming vehicles. VANET is a high mobile network where the nodes are moving in speeds that may exceed 120km/h, which means that this vehicle move 33.33m/s, even if these vehicles are very far from the danger, they will reach it very soon, here milliseconds will be very important to avoid the danger [6, and 7].

Emergency messages in VANET are sent in broadcast fashion where all the vehicle inside the coverage area of the sender should receive the message. The coverage area is not enough as it is hardly reaches a 1000m (which is the DSRC communication range) due to attenuation and fading effects. Away vehicles from the danger should receive this critical information to avoid the danger. Furthermore, the probability of message reception can reach 99% in short distances and can be as low as 20% at half of the communication range (Moreno, 2004). Therefore, there should be a technique to increase the emergency message reception with high reliability and availability.

Duo to the high mobility of vehicles, the distribution of nodes within the network changes rapidly, and unexpectedly that wireless links initialize and break down frequently and unpredictably. Therefore, broadcasting of messages in VANETs plays a crucial rule in almost every application and requires novel solutions that are different from any other form of Ad-Hoc networks. Broadcasting of messages in VANETs is still an open research challenge and needs some efforts to reach an optimum solution.

Broadcasting requirements are: high reliability and high dissemination speed with short latency in single-hop as well as multi-hop communications. Problems associated with regular broadcasting algorithms are: the high probability of collision in the broadcasted messages, the lack of feedback and the hidden node problem.

In this paper we concerned with proposing a new intelligent broadcasting technique for the emergency message in VANET aiming to increase the reception of the emergency information.

## II. Research Background

Emergency message rebroadcast

In [8], authors proposed a street-based broadcast scheme that utilizes neighbor's information by exchanging hello messages among vehicles, when any probable danger is detected, a warning message is broadcasted to all neighbors. The farthest vehicle is selected as a forwarder depending on the information gained from the hello message, if the preselected forwarder receives the message, it will rebroadcast it.

Depending on just one forwarder is not enough in a high mobile network like VANET. Furthermore, authors didn't depend on beacons to gain the information. They proposed to use hello message, which creates a chance to increase the channel load.

The contention period schemes (which is a waiting time that the receiver waits before rebroadcasting the original message received from the sender) are proposed by many researchers[9], [10], [11], [12], [13], [14] and [15].

In [9], authors proposed the Link-based Distributed Multi-hop Broadcast (LDMB), in which all the receivers of the emergency message are potential forwarders. Each forwarder computes and waits for contention time using equation (3.2), if the contention time ends the forwarder will start to rebroadcast the emergency message.

In [11] and [12], where authors proposed position-based message forwarding strategy by sending the emergency message in a broadcast fashion, and selecting the best forwarder available. All vehicles receiving that message are potential forwarders. In order to decide which node forwards the message all receivers will be assigned a contention window (waiting time); the contention window size will be the smallest for the farthest node and the biggest size for the nearest node, in other words, this protocol will give priority for the farthest node to be the next forwarder.

The problem of the last two protocols that all the message receivers will compute the waiting time and wait to make the rebroadcast even the closest vehicles to the sender will do and this will make the entire network vehicles busy for any message received.

Another protocol proposed by [13] called Emergency Message Dissemination for Vehicular (EMDV) protocol, by enabling the farthest vehicle within the transmission range to make the rebroadcasting of the emergency message.

Choosing one forwarder vehicle is not appropriate in a high mobile network like VANET as the position is always changing, and the receiver vehicle may become out of range when sending the message or simply the receiver can't receive the message because of the channel problems like jam or denial of service, see figure 2.

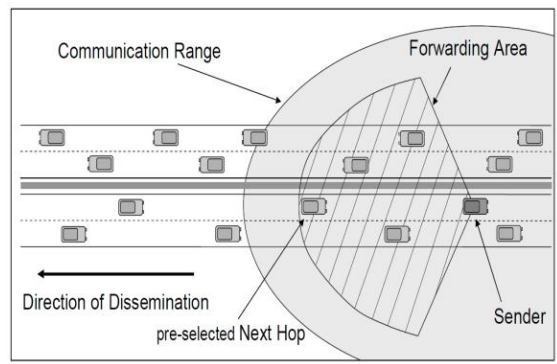

Figure 2: Sender utilizing EMDV

In [10] authors proposed that the receivers of the message will select random waiting times and make acknowledgment to avoid the re-transmissions from nodes closer to the original sender.

The acknowledgment scheme causes delay to the rebroadcast.

In [14] authors proposed the Contention-Based Forwarding (CBF) protocol where a vehicle sends a packet as a broadcast message to all its neighbors. On receiving the packet, neighboring vehicle will contend for forwarding the packet. The node having the maximum progress to the destination will have the shortest contention time and will first rebroadcast the packet. If other nodes receive the rebroadcast message, they will stop their contention and delete the previously received message. This protocol mainly proposed for forwarding the periodic safety message (Beacons).

The problem of this protocol that there should be a management technique to manage the contention for all the neighboring vehicles, and there is a chance that the nearest vehicle to the sender may not hear the rebroadcast of another vehicle, here this vehicle will rebroadcast the message and this called (hidden node problem (Tobagi and Kleinrock, 1975)) also it may lead to broadcast storm problem that makes the protocol useless.

In [15] authors suggested that the emergency message will be rebroadcasted by the receivers located at farther distances from the sender by the selection of shorter waiting times, see equation 1.

In [7] authors proposed the Contention Based Broadcasting (CBB) protocol for increasing the emergency message reception and performance, the emergency message will be broadcasted in multi-hop fashion, and the multi-hop forwarders will be selected before the original message is sent. CBB proven to achieve superiority over the EMDV protocol as it choses more than one forwarder to rebroadcast the emergency information and this gives the message a chance to overcome the preselected forwarder failure.

The criteria of choosing the forwarders depends on the progress and on the segment localization, see figure 3, where all the vehicles located in the final; none-empty segment are a potential forwarder.

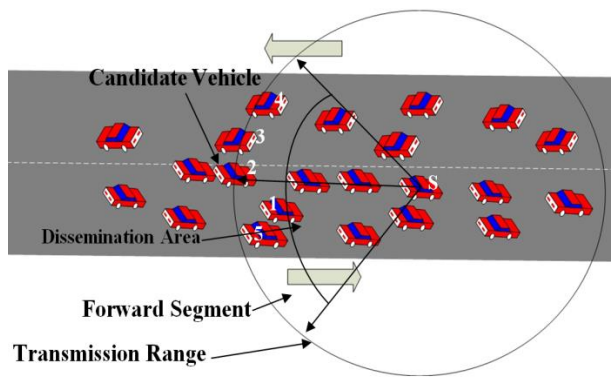

Figure 3: Emergency message Sending and transmission range

*Emergency message rebroadcast by network segments*

Another way to rebroadcast the message is to divide the network into segments proposed in [16, 17, and 18].

In [16] authors proposed a protocol called Urban Multi–hop broadcast (UMB) aiming to maximize the message progress, and avoid broadcast storm, hidden node, and reliability problems. The protocol assigns the duty of forwarding and acknowledging the broadcast packets to only one vehicle by dividing the road portion inside the transmission range into segments and choosing the vehicle in the furthest non-empty segment without prior topology information. The source node transmits a broadcast control packet, called Request to Broadcast (RTB), which contains the position of the source and the segment size. On receiving the RTB packet, nodes compute the distance between the sender and the receiver. Then, nodes transmit a channel jamming signal, called black–burst, that contains several time–slots equal to their distance from the source (in number of segments): the farther the distance, the longer the black–burst. Each node transmits its black-burst and senses the channel; if there is no other black-burst in the channel it concludes that it is the farthest node from the source. Then the node returns a Clear–to–Broadcast (CTB) control packet, containing its identifier (ID), to the source.

The Smart Broadcasting Protocol [17] addressed the same objective as UMB using a different methodology. Upon reception of a RTB message, each vehicle should determine its segment and set a random back-off time. Each segment has its own contention window size, i.e. if this segment has contention window size (4) TS (time-slot); vehicles in the furthest segment should randomly choose a back-off time between (0) to (3) TS. Vehicles in the next nearer segment choose a value between (4) to (7) TS, and so on, as vehicles near the sender should wait for longer time.

Vehicles will decrement their backoff timers by one in each time-slot while listening to the physical channel. While waiting, if any vehicle receives a valid CTB message, it will exit the contention phase and listen to the incoming broadcast. On the contrary, if any node finishes its backoff timer, it will send the CTB containing its identity and rebroadcast any incoming broadcast.

While in [18] authors proposed the Geographic random forwarding (GeRaF) protocol, which divides the network into equally adjacent sectors, the transmitter (source) elects the sectors starting from the farthest one, by sending RTB message, all the nodes in the elected sectors reply by CTB message, if one node reply the CTB message, then this node will become the next forwarder, if there are more than one node sent the CTB message the source issue a collision message and make a collision-resolution procedure to elect the next forwarder depending on a probabilistic rule.

Many other approaches are discussed in details in our previous paper (emergency message broadcasting).

### III. THE PROPOSED PROTOCOL

This section presents a detailed design description for the PCBB protocol, which aims to increase the percentage of reception for the emergency information by utilizing a contention window, position based forwarding scheme, and PSO intelligent technique.

Beacons and the emergency messages should be received by all the neighboring vehicles with high probability and reliability, because of the critical nature of the information both provide. When a vehicle detects danger, it issues an emergency message to warn other vehicles within the network, and all the vehicles in the opposite direction of the sender movement located in its transmission range must receive such message. Covering the whole area does not guarantee that all vehicles will receive the message because of channel collisions and fading effects. The percentage of emergency message reception for the network vehicles must be as high as possible.

In order to cover a wider area for message reception, some neighboring vehicles can serve as potential forwarders, and each forwarder has to wait for a certain period of time (i.e., contention time) before forwarding the message.

*Preparing to send*

Every beacon received by a vehicle provides important information about the sender status. This information is utilized to form a rich and real time image about the current network topology, which facilitates better network vehicle communication. It also helps to be informed about the potential dangers when they occur. When a vehicle has a problem or detects a problem, it determines if the problem is life critical or not. The life critical (Safety of Life) messages will be given the highest priority and are then processed and sent before any other kind of messages.

This paper proposes to categorize any emergency message before sending it to make it easier for the receiver vehicle to recognize the importance of the message being received. Table 1 lists the codes for each category. For example, when a vehicle receives two messages containing categories 1 and 5, it processes the message that contains category 1 first because it contains more critical information.

After assigning the message code, the sender should add data to the message, such as the coordinates of the danger zone or what the receiver should do; however, this aspect would not be discussed in detail in the current study. The proposed structure of the message is shown in Figure 4, where three inputs, namely *Cid*, *MinB* and *MaxB*, are added to help the receiver vehicle determine what action to take after receiving the emergency message.

TABLE 1: EMERGENCY MESSAGE CLASSIFICATION

| Code | Priority | Application |
|------|----------|-------------|
| 001 | Safety of Life | Emergency Break Warning/Avoidance |
| 002 | Safety of Life | Cooperative Collision Warning |
| 003 | Safety | Intersection warning. |
| 004 | Safety | Transit Vehicle Signal Priority |
| 005 | Non-Safety | Toll Collection |
| 006 | Non-Safety | Service Announcement |
| 007 | Non-Safety | Movie Download(2 hours of MPEG 1) |

Sending the message in single hop enables it to reach a number of vehicles within a limited distance up to 1000 m for the best cases [19]. However, this number should be increased in order to warn more vehicles of possible dangers before they reach the danger area.

| Sen ID | Code | TS | Msg ID | Data | CId | MinB | MaxB |
|--------|------|-----|--------|------|-----|------|------|

Figure 4: Emergency message illustration.

Where Sen ID: sender id, Code: Message Code, TS: Time Stamp, Msg ID: Message ID, Data: Data sent, CId: Forwarder Candidate ID, MinB: Minimum Boundary, MaxB: Maximum Boundary.

Choosing the next candidate forwarder is a process, which begins by gathering the information obtained from beacons received from neighbors. This information is inserted and ordered into NT. The sender vehicle chooses the farthest vehicle and assigns it to be the candidate forwarder. The process of forwarding the emergency message to increase the probability of reception so that the forwarded signal can communicate with more vehicles on the road and reach longer distances is the option used in this paper. Choosing only one forwarder is inappropriate in high mobile networks, such as VANET, because the forwarder might not receive the emergency message. To solve this problem, dividing the network to several segments is proposed. Vehicles inside the last non-empty segment (i.e., the farthest segment from the sender) wait for a period of time and determine whether or not the candidate forwarder rebroadcasted the emergency message. If none made the rebroadcast, the vehicles located in the farthest segment forwards the message. As mentioned earlier, assigning the forwarding job of the emergency message to all the receiver vehicles of the message may cause a broadcast storm problem, and assigning the forwarding job to just one receiver vehicle may be inappropriate because the specific forwarder may not receive the emergency message. Hence, vehicles in the last segment must forward the emergency message if the forwarder fails to receive and forward the message.

As mentioned earlier assigning the forwarding job of the emergency message to all the receiver vehicles of the message may cause a broadcast storm problem, and assigning the forwarding job just for one receiver vehicle may not be appropriate sometimes as this specific forwarder may not receive the emergency message. Hence vehicles in the last segment, which should be the furthest one, will make the forwarding of the emergency message if the forwarder fails to receive and forward the message.

As mentioned earlier, the network is divided to several segments to help the vehicle determine the next forwarder of the emergency message. As proposed in [16], the transmission range of the sender is divided into 10 segments to make it easier for the sender to determine the last vehicle in the last non-empty segment, which is eventually selected as the next forwarder. For this paper, the distance is between the sender and the forwarder. Authors in [16] established a fixed distance of 1000 m. For the current study, however, if the distance between the sender and the farthest vehicle is 900 m, anything beyond 900 m is not considered. The distance between the last vehicle and the sender is computed using Equation (1).

$$Dis = SenPos - ForPos \qquad (1)$$

where *Dis* is the distance between last vehicle and the sender, *SenPos* is the position of the sender obtained from GPS, and *ForPos* is the forwarder position or the last vehicle in the last non-empty segment.

Determining the boundaries of the last non-empty segment must be set dynamically depending on the channel status and the network topology available, because it would be pointless if this segment does not contain enough number of vehicles for forwarding. At the same time, determining the number of sufficient vehicles located in the last non-empty segment must also depend on the channel status and network topology. The sender vehicle has all the information required to analyze the channel and draw the network topology. To compute the boundaries, the CBB and PCBB protocols are proposed. The CBB protocol depends on the selection of boundaries of the last non-empty segment based on the number of the vehicles located in this segment, and the number of segments in the network. [16] suggested that the number of the segments should be 10 segments. Computing the segments and the boundaries could be done using Equations (2), (3) and (4). Equation (2) assigns the distance between the sender, and the farthest vehicle (forwarder) is the boundary of the last non-empty segment. Equation (3) computes the length of each segment, and Equation (4) finds the location of the minimum boundary. *MinB* is the minimum boundaries (Borders) where the last segment starts. *Nmax* is the maximum number of the segments, and *Dif* is the length of the segment.

$$MaxB = Dis \qquad (2)$$

$$Dif = \frac{Dis}{Nmax} \qquad (3)$$

$$MinB = Dist - Dif \qquad (4)$$

This means that the vehicles located in the area between *MinB* and *MaxB* from the sender are considered as potential forwarders of the emergency message. They are rebroadcast if no vehicle makes the rebroadcast. Sometimes, the last segment may have an insufficient number of vehicles. The number of the potential forwarders must have a threshold to determine if it is sufficient or not.

$$SucPer = \left(\frac{NeiN}{Nmax} - 1\right) \times 100\% \qquad (5)$$

The number could be generated and tested using Equation (5), where *SucPer* is the success percentage that the last segment must fulfill before agreeing on the values of *MaxB* and *MinB*, and *NeiN* is the total number of neighbor vehicles in *NT*.

If *SucPer > Nmax%*, this means that the last segment holds enough number of potential forwarder vehicles. The result of $\frac{NeiN}{Nmax}$ is subtracted by one vehicle, because the last segment also holds the preselected potential forwarder. If *SucPer < Nmax%*, this means that the area of the last segment should be expanded to include more vehicles, and this could be determined using Equation (6).

$$MinB = Dist - 2 \times Dif \qquad (6)$$

This calculation doubles the size of the last segment and increases the number of the potential forwarders. If the calculated number remains to be *SucPer < Nmax%*, the *MinB* could be recalculated by multiplying *Dif* by 3 and so on. This technique increases the number of the potential forwarders and solves the preselected forwarder rebroadcast failure.

The PCBB is an enhancement of the CBB and works when the sender vehicle analyzes the dense locations of the vehicles along its transmission range. In Figure 5, the vehicle analyzes the location density in the network to form groups for dense locations; the resulting network is then divided into several groups (Figure 6).

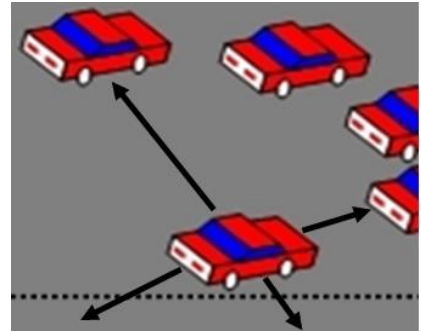

Figure 5: vehicle analysis location density.

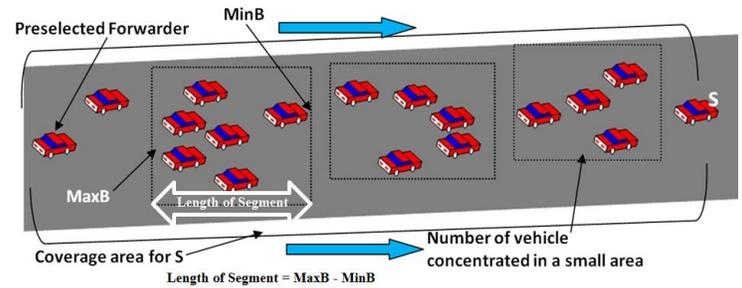

Figure 6: Analyzed network depending on network density and progress.

A dense or concentrated area has a number of vehicles within just a small area. Thus, sending a message to vehicles in concentrated areas increases the chance of receiving and rebroadcasting the message. The probability of receiving the rebroadcasted messages in this segment is also high, thus eliminating the hidden node problem, which is considered one of the most difficult problems encountered in rebroadcasting emergency message in VANET. Equation (7) is used to calculate how vehicles compute the dense locations. The progress represents the upper bound of the last segment and the length of the segment is also the distance between the farthest vehicle in the segment and the first vehicle located in the segment. For example, if the segment that has progressed to 820 m, has 15 vehicles in 80 meters between 820 m and 900 m from the location of the sender, the segments with the higher progress and high number of vehicles with smaller segments give higher fitness function.

$$Fitness = \frac{Progress \times No.\,of\,Vehicles}{Lenght\,of\,Segment} \quad (7)$$

After performing Equation (7), the vehicle inserts it in a Progress List (PL), which then helps the vehicle in making quicker analysis and decisions (Table 2).

TABLE 2: PROGRESS LIST.

| Progress (m) | Length of Segment | No. of Vehicles | Fitness Function |
|---|---|---|---|
| 520 | 90 | 15 | 86 |
| 610 | 90 | 15 | 101 |
| 700 | 120 | 30 | 175 |
| 820 | 80 | 15 | 153 |
| 900 | 0 | 1 | 0 |

After performing Equation (7) on all the vehicles in the segments, the sender vehicle takes the upper boundary of the segment scores the higher fitness function then the PSO optimization is applied

$Fit_v = lBest_v * w + C1 * rand1 * (pBest_v - lBest_v) + C2$

$* rand2 * (gBest_v - lBest_v),$ (8) [20].

$lBest_v = pBest_v + Fit_v$ (9), [20].

Where *W* is random number between *W*: 0.1 to 0.5, *C1*= 2, *C2*= 2, rand: random number 0.1 to 1, *pBest* is the last *lBest* computed by the vehicle. *w* is the inertia weight of the particles, random 1 and random 2 are two uniformly distributed random numbers in the range [0, 1], and *C1* and *C2* are specific parameters which control the relative effect of the individual and global best particles.

The *lBest* for the vehicle is obtained from the fitness function computed using (7), which represents the best area (segment dimension). The results indicate the sufficient number of vehicles depending on the sender analysis. *Pbest* is the previous fitness function computed by the vehicle, while the *gBest* is best fitness function computed by the vehicle obtained from the analysis of the information from CRNT obtained from our previous paper [7]. The CRNT gives extended information received from other vehicles located in the neighborhood of the sender, which reduces the error possibility that the vehicle might make during the channel dense location analysis because PSO depends on taking the neighbor's information and history. From the CRNT, the vehicle can conclude another global fitness function from the neighboring vehicles analysis, which influences the current analysis done by the current vehicle

To compute for the boundaries of the last segment, the sender carries out the following equations:

$MaxB = Dis$ (10)

$MinB = lBest$ (11)

Where *MaxB* is the highest boundary (border) for the last segment, and *MinB* is the minimum boundary (Borders) from which the segment starts.

This means that the vehicles located in the area between *MinB* and *MaxB* from the sender are considered as potential forwarders of the emergency message, and would have to wait to rebroadcast in case no vehicle forwards the message.

When the sender decides to broadcast an emergency message, it should examine the number of neighbors within the back end of the coverage area. If the number is more than one, then this protocol could be carried out; if the number of the neighbors is zero, then the sender broadcasts the message without specifying any forwarder. If the number of the neighbors is equal to one, then the sender broadcasts the message and specifies the forwarder without adding any detail about the boundaries.

To compute for the contention time, Equation (12) is performed, where the vehicles with the largest distance from the sender have the shortest contention time to wait before testing the channel to rebroadcast. Each vehicle tests its progress from the sender by dividing its current position on the maximum distance computed by the sender. The result of this equation gives the waiting time for the contending vehicles inside the last segment, giving the opportunity for all the vehicles inside the last segment to recover from the failure of the chosen forwarder. Thus, the protocol increases the probability of resending the emergency message consequently, increasing the percentage of sending the emergency message and reaching longer distances at the same time.

Enabling just the last segment to contend eliminates the hidden node problem, because all the potential forwarders have high probabilities to sense the rebroadcasted message when a forwarder resends the message. All potential forwarders are located in a small and limited area. The probability of reception can reach 99% at short distances, but it could be as low as 20% at half the distance of the communication range [13].

Sending Steps: The sender dispatches an emergency message warning other vehicles about any potential danger. The sender analyzes the danger and selects the code. The sender creates NT for its neighbors, and then selects the next forwarder depending on the distance of the farthest vehicle. The sender analyzes the dense location computing the fitness function using Equation (7). The sender analyzes the information gained from neighbors about dense locations from CRNT and concludes the *gBest*. The PSO algorithm is then applied to obtain the *MinB*, which represents the lower bound of the segment. The sender creates the message and inserts the values derived from steps 4 and 8 in the message, after which it broadcasts the message to the network

The following illustrates the calculations using Equation (7), which computes the fitness function for each segment. This formula ensures that the vehicles having high progress from the sender and having a large number of vehicles in a small area can produce better fitness function. This is because vehicles concentrated in a small area and are located far from the sender vehicle have better opportunities to rebroadcast the emergency message with little chance of failure.

From Table 2 and after employing Equation (7):

$$Fitness\ Function = \frac{700\ x\ 30}{120} = 175$$

The best value for the fitness function is 175; this value means that the *lBest* is 700, which represents the lower boundary for that segment.

The *gBest* is taken from the CRNT, because this protocol provides the sender vehicle with information from other vehicles. This information also enables the sender to analyze the channel depending on the information from the other vehicle giving more accurate data about the network.

*pBest* is the channel analysis history of the network made from the sender vehicle, *lBest* is the boundary of the fitness function, and gBest is the best analysis from the neighbors.

lBest = 700, pBest= 690, gBest= 720.

Apply PSO equation (8)

700 x 0.1 + 2 x 0.65(690 − 700) + 2 x 0.7 (720 − 700) = 70 − 13 + 28 =95

lBest = 690 + 95 eq (4.10)

lBest = 785.

The *MinB* for this example is 785 m and the *MaxB* is 900 m, as obtained from Equation (10). The results imply that the vehicles between 785 m and 900 m from the sender can be considered as potential forwarders; they would have to contend by means of the time to rebroadcast and be able to overcome the preselected forwarder failure.

### Receiving a message

Vehicles in VANET receive messages all the time. These messages could be beacon, emergency, or service messages. Each message should be analyzed to determine the level of importance involved. If the message is holding safety critical information, then the message code should be 1 or 2, and the message is given higher priority for processing. The receiver vehicle also has to ensure that this message has not been received before to eliminate duplication. The receiver checks the forwarder *ID*. If the current receiver *ID* is the same as that of the forwarder, the receiver rebroadcasts the message immediately. If the receiver is not a forwarder, then it must compute the distance between its position and that of the sender to determine if the receiver is located within the last non-empty segment. If the receiver is located in the last non-empty segment, it starts contending and prepares itself to rebroadcast. Each vehicle inside the last non-empty segment must contend the time and compute the *CW* time (Waiting time) using Equation (12). *CW* depends on the progress, and the vehicles having the largest progress from the sender has the shortest contention time before testing the channel to make the rebroadcast.

$$Tc = Tslot \times \left( \left( 1 - \frac{Dis}{MaxB} \right) \times 100 \right)\ (12)$$

*Tc* is the contention time that the segment vehicle has to wait before checking the system to see if the emergency message has been rebroadcasted by other vehicle or not, *Tslot*: is the system time slot.

Receiving emergency message steps: This section represents the steps for receiving the emergency message, which must be done efficiently. The receiver accepts a message then checks the code if it is 1 or 2. The receiver also checks if the message has been received before and if its *ID* is the same as the forwarder *ID*, then, it rebroadcasts the message immediately. If the *ID* is not the same, the receiver calculates the distance between its current position and the sender and tests if the current location falls within *MinB* and *MaxB*. The receiver then prepares to forward the message.

Rebroadcasting steps:

The rebroadcasting job is only assigned to a limited number of vehicles. It is not appropriate to assign this job to all the receiver vehicles, because this can lead to a broadcast storm problem [22] or the hidden node problem. The first forwarder should be the first candidate selected by the sender, and the forwarding steps are as follows.

The forwarder waits for a random back off time depending on its contention window. The back off time is used to avoid channel collision. The forwarder then senses the channel and tests whether or not this message has been transmitted from others. If no other vehicle rebroadcasts, the message forwarder reserves the channel and the forwarder broadcasts the message. The contention window for the first candidate forwarder is 15 µs.

Each vehicle inside the last non-empty segment is required to contend before trying to send the emergency message and this could be done using the following steps. The forwarder computes the contention window using Equation (12). The vehicle contends for a random back off time depending on the contention window. The vehicle senses the channel and determines whether or not this message has been transmitted from others. If no other vehicle rebroadcasts the message, the vehicle reserves the channel and broadcasts the message.

PCBB and CBB both have the same goal of increasing the percentage reception of the emergency information. The main difference between them, however, is the potential forwarder selection, where CBB depends on choosing the last non-empty segment boundaries depending on the number of the vehicles in the segment against a predefined threshold, whereas the PCBB depends on selecting the boundaries on vehicle saturation areas and utilizes the PSO intelligent technique.

### IV. SIMULATION

Simulation Setup

In order to test correctness of our protocol we made the simulation using the commercial program Matlab®, the distribution used is Nakagami distribution.

Parameters used in our simulation are summarized in table 1; all the simulations in this paper will adopt these parameters.

We made our simulation for 10s including 200 vehicles in 2 km road consisting of 3 lanes.

Simulation Parameters

Parameters used in the simulation experiment are summarized in table 3; all the simulations in this paper will adopt these parameters.

<p style="text-align:center">TABLE 3: SIMULATION CONFIGURATION PARAMETERS</p>

| Parameter | Value | Description |
|---|---|---|
| Radio propagation model | Nakagami-m, m = 3 | Model m=3 is fixed value, recommended by [13] |
| IEEE 802.11p data rate | 6Mbps | Fixed value |
| PLCP header length | 8 µs | Fixed value |
| Symbol duration | 8 µs | Fixed value |
| Noise floor | -99dBm | Fixed value |
| SNR | 10 - 40 dB | Adjustable to add noise to the signal |
| CW Min | 15 µs | Fixed value |
| CW Max | 1023 µs | Fixed value |
| Slot time | 16 µs | Fixed value |
| SIFS time | 32 µs | Fixed value |
| DIFS time | 64 µs | Fixed value |
| Message size | 512 bytes | Fixed value |
| Beacon Message Rate | 10 Message / s | Fixed value |
| Number of Vehicles | 200 | Fixed value |
| Road Length | 2 KM | Fixed value |
| Car Speed | 20km – 120km | Fixed value |
| Simulation Time | 10 s | Fixed value |
| Road Type | Highway | Fixed value |
| Number of lanes | 3 lanes | Fixed value |
| Neighbor entry size | 15 Bytes | Fixed value |

## V. RESULTS

In order to enhance emergency message dissemination in VANET, two contention- and position-based protocols have been proposed and implemented, namely, CBB and PCBB (an enhancement of the proposed CBB). In this section, the EMDV protocol, which is the outcome of the NOW project [13, and 22], is compared with the CBB and PCBB. The results of the experiment are shown in Figures 7, 8, and 9.

The test performed concentrated on the probability of emergency message reception, channel collision, and the delay that the protocols may cause. The EMDV and DFPAV protocols, both widely used in VANET today, are the results of the NOW project, which is a collaboration between Mercedes-Benz and Karlsruhe University [22]. Figure 7 shows the simulation results for the proposed CBB and PCBB protocols. These have been simulated and tested in terms of probability of emergency message reception; afterwards, their performances are compared with that of the EMDV protocol. The results show that all the protocols can increase the performance and probability of emergency message reception; more noteworthy is the fact that CBB and

PCBB can achieve better performance past the first 1000 m distance representing the DSRC communication range for the sender. The signal gets very weak in the last meters of the sender communication range, but when the forwarder rebroadcasts the message, the signal becomes stronger and reaches greater distances.

Another difference between CBB/PCBB and EMDV is that after several tries, CBB and PCBB never fail to rebroadcast the emergency message, but EMDV sometimes fails to do so, proving the effectiveness of the CBB protocol. The PCBB protocol has also been shown to select forwarders more carefully than CBB, which depends on a threshold, while PCBB depends on traffic saturation and progress and on the analysis made by neighboring vehicles. Furthermore, PSO adopts the PSO intelligent algorithm, which takes other vehicles' analysis into consideration, allowing for more accurate selection of preselected forwarders.

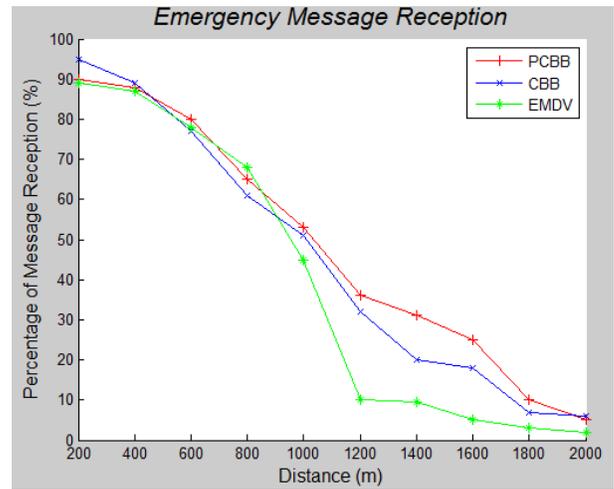

Figure 7: Probability of message reception of emergency message with respect to the distance to the sender.

Figure 8 shows the message delay for CBB and EMDV compared with PCBB emergency message delay. The simulation computes the delay for broadcasting and rebroadcasting of the original message, showing that the EMDV during the time has a slightly higher delay than CBB but not exceeding 50 µs. The delay shows a slight increase at about 20 µs 1000 m away from the sender where the rebroadcast starts to take effect. If the CBB has a shorter delay starting from this point, it means that its rebroadcast efficiency and decisions are made faster than those in EMDV. PCBB has a slightly shorter delay (about 10 µs shorter than CBB and 30 µs shorter than EMDV at the ninth second), because the PSO is an intelligent technique that has quick performance and response [23]. In safety systems with a highly mobile network like VANET, a few microseconds are critical in saving life or avoiding danger.

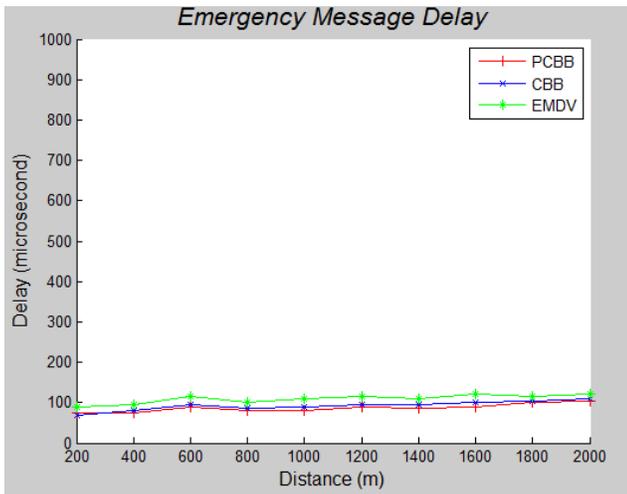

Figure 8: Delay measured after sending the emergency message with respect to distance.

Figure 9 shows the collision produced by the three protocols, all of which generated the same collision when broadcasting emergency information. It is worth noting that the collisions produced by CBB, PCBB and EMDV at the beginning of the experiment do not increase. However, after a period of time, sending a large number of emergency messages resulted in an increase in the number of collision for all the three protocols, with the difference between them reaching 1% at the ninth second.

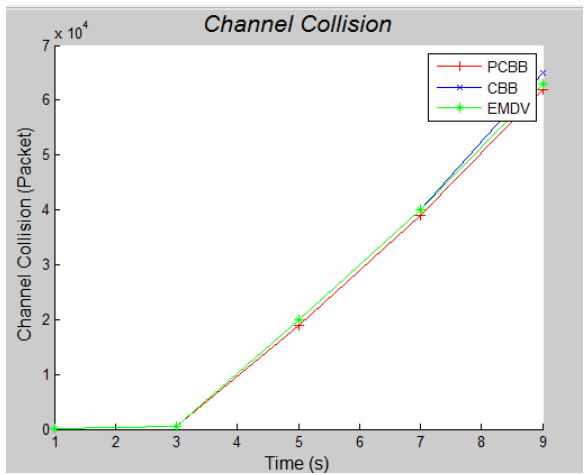

Figure 9: Collision measured after sending the emergency message.

## VI. CONCLUSION

This Research has proposed the PCBB aiming to improve road safety by achieving fast and efficient emergency message transmission and delivery, utilizing the efficient and newest intelligent technique (PSO), which helped to make more accurate analysis and performance, and increased the percentage of the emergency message reception without affecting the channel collision.

## REFERENCES

[1] Ghassan Samara, Wafaa AH Al-Salihy, R Sures, "Security Issues and Challenges of Vehicular Ad Hoc Networks (VANET)", 4th International Conference on New Trends in Information Science and Service Science (NISS), 2010, 393-398.

[2] Ghassan Samara, Wafaa AH Al-Salihy, R Sures, "Security Analysis of Vehicular Ad Hoc Nerworks (VANET)", Second International Conference on Network Applications Protocols and Services (NETAPPS), 2010, 55-60.

[3] Who, W. H. (2011), World Health Organization, http://www.who.int/roadsafety/decade_of_action/plan/en/index.html, visited on 30 April, 2014.

[4] M Raya, D. J., P Papadimitratos, I Aad, JP Hubaux (2006), Certificate Revocation in Vehicular Networks. Laboratory for computer Communications and Applications (LCA) School of Computer and Communication Sciences, EPFL, Switzerland.

[5] Worldometers (2011), real time world statistics, www.worldometers.info, visited on 30 April, 2014.

[6] Ghassan Samara, WAHA Alsalihy, S Ramadass, "Increase Emergency Message Reception in VANET", Journal of Applied Sciences, 2011, Volume 11, Pages 2606-2612.

[7] Ghassan Samara, Wafaa Alsalihy, Sureswaran Ramadass," Increasing Network Visibility Using Coded Repitition Beacon Piggybacking", World Applied Sciences Journal (WASJ), Volume 13, Number 1, pp. 100 - 108, 2011.

[8] Ching-Yi, Y. & Shou-Chih, L. (2010) Street Broadcast with Smart Relay for Emergency Messages in VANET. 24th International Conference on Advanced Information Networking and Applications Workshops (WAINA), 323-328, IEEE.

[9] Qiong, Y. & Lianfeng, S. (2010) A Multi-Hop Broadcast scheme for propagation of emergency messages in VANET. 12th IEEE International Conference on Communication Technology (ICCT), 1072-1075, IEEE.

[10] Biswas, S., Tatchikou, R., Dion, F. (2006), Vehicle-to-vehicle wireless communication protocols for enhancing highway traffic safety, IEEE Communications Magazine, 44, 74-82, IEEE.

[11] Torrent-Moreno, M. (2007), Inter-vehicle communications: assessing information dissemination under safety constraints, 4th Annual Conference on Wireless on Demand Network Systems and Services, WONS '07, 59 – 64, IEEE.

[12] Torrent-Moreno, M., Mittag, J., Santi, P. & Hartenstein, H. (2009), Vehicle-to-Vehicle Communication: Fair transmit power control for safety-critical information. Transactions on Vehicular Technology, 58, 3684-3703, IEEE

[13] Torrent-Moreno, M. (2007), Inter-Vehicle Communications: Achieving Safety in a Distributed Wireless Environment-Challenges, Systems and Protocols, (Ph.D Paper), Universitatsverlag Karlsruhe, ISBN: 978-3-86644-175-0.

[14] Füßler, H., Widmer, J., Kasemann, M., Mauve, M. & Hartenstein, H. (2003), Contention-based forwarding for mobile ad hoc networks, Ad Hoc Networks, 1 (2003), 351-369.

[15] Briesemeister, L., Schafers, L., Hommel, G. (2000), Disseminating messages among highly mobile hosts based on inter-vehicle communication, IEEE Intelligent Vehicles Symposium, IV 2000, 522 - 527, IEEE.

[16] Korkmaz, G., Ekici, E., Özgüner, F. & Özgüner, Ü. (2004), Urban multi-hop broadcast protocol for inter-vehicle communication systems, 1st ACM international workshop on Vehicular ad hoc networks, ACM.

[17] Fasolo, E., Zanella, A., Zorzi, M. (2006), An effective broadcast scheme for alert message propagation in vehicular ad hoc networks, IEEE Int. Conf. on Communications ICC'06, 3960 – 3965, IEEE.

[18] Zorzi, M. & Rao, R. R. (2003), Geographic random forwarding (GeRaF) for ad hoc and sensor networks: energy and latency performance, IEEE transactions on Mobile Computing, 349-365, IEEE.

[19] IEEE (2005), White Paper: DSRC technology and the DSRC Industry consortium (DIC) prototype team.

[20] Neo Project, (2014), http://neo.lcc.uma.es/staff/jamal/portal/?q=content/particle-swarm-optimization-pso, Malaga University, visited on 10 March 2014.

[21] Ni, S. Y., Tseng, Y. C., Chen, Y. S. & Sheu, J. P. (1999a), The broadcast storm problem in a mobile ad hoc network, 5th annual ACM/IEEE international conference on Mobile computing and networking, 151-162, ACM.

[22] Now, (2011). Network on Wheels project, http://www.network-on-wheels.de/vision.html, accessed 5 May 2011.

[23] Mendes, R. (2004). Population Topologies and Their Influence in Particle Swarm Performance (PhD thesis). Universidade do Minho.

APPENDIX

---

**Procedure** DetectDanger ( )
{
Gather neighbor information
Select main forwarder
MaxB = forwarderlocation – senderlocation.
(MinB) = PSO ( )
Insert Emergency information in the message
Send Message
} % end procedure

**Procedure** PSO ( )
{
Currentsegment = Computevehicleconcentration ( ) % to compute the current concentration of vehicles
pBest = cfitness
cfitness = 0
For i=0 to Currentsegment size -1 % calculate the best result

$$Fitness = \frac{Segment\ (idist) \times Segment\ (iprogreaa)}{Segemnt\ (i\ Segment\ Vehicles)}$$ % fitness function

If Fitness > cfitness
cfitness = Fitness
} % end if
} % end for
lBest = cfitness

neighborsegment = Computevehicleconcentration ( ) % to compute the current concentration of vehicles
gfitness = 0
For x=0 to Currentsegment size -1 % calculate the best result
{

$$Fitness = \frac{Segment\ (x,dist) \times Segment\ (x,progress)}{Segemnt\ (x\ Segment\ Vehicles)}$$ % fitness function

If Fitness > cfitness
gfitness = Fitness
} % end if
} % end for
gBest = gfitness

}% end procedure

**Procedure** ReceiveEmerMessage ( )
{
If code = 1 or code = 2
{
If preselctedforwarder
{
Rebroadcast ( )
Else if candidate forwarder
{
      Compute CW
      Choose random backoff
      While back <> 0
        {

---

Wait contention time
If channel = idle
{
    If no rebroadcast
    {
      Rebroadcast ( )
    } % end if

    Else
    {
    Backoff = backoff -1
    } % end if
    } % end while
} % end if
} % end if
} % end if
}% end procedure

**Procedure** Rebroadcast
Rebroadcast emergency message
End % procedure

Procedure Computevehicleconcentration ( )
{
Mindist = Computemindist ( )
Arrange NT descending
For i= 0 NT size -1
{
Take the average between two successive vehicles
If average < 2 * (last average) or number of vehicles compared is 1 % if this vehicle is the first vehicle
{
Add to current segment % add the current vehicle to this segment
Segment vehicles = segment vehicles + 1 % increase the number of vehicles by 1
Vehicle location = location (i) %takes the location of the current vehicle
Segment (segment count, dist) = Vehicle location - first element location % compute the width of the

segment
segment (segment count, Segment vehicles) = Segment vehicles % to store the number of vehicles for the

current segment

}
Else if the new average is more than the double of previous value
{
segment count = segment count + 1
Vehicle location = location (i)
Segment vehicles = 1
Segment (segment count, progress) = Vehicle location
}
**Return** (segment)
}

---

Figure A.1: Psedu-code Particle Swarm Optimization Contention Based Broadcast Protocol (PCBB).

The procedure DetectDanger works when the vehicle detects any danger, the first step for the sender is to order the neighbors' information in NT, and select the first forwarder, afterwards. It calls the PSO procedure that implements the PSO algorithm to select the vehicles that overcome the preselected forwarder's failure.

The procedure ReceiveEmerMessage works when the vehicle receives an emergency message and checks if the receiver is a forwarder or not.

The procedure Computevehicleconcentration analyzes the neighbors to discover the location of the vehicles' concentration.